\documentclass[aps,prl,reprint,superscriptaddress]{revtex4-2}
\usepackage{graphicx}
\usepackage{subfigure}
\usepackage{epsfig}
\usepackage{xcolor}
\usepackage{dcolumn}
\usepackage{bm}
\usepackage{color}
\usepackage{amsmath}
\usepackage{amsfonts}
\usepackage{amssymb}
\usepackage{float}
\usepackage{epstopdf}
\usepackage{caption3}
\usepackage{epstopdf}
\usepackage{multirow}
\usepackage{longtable}

\def\Eq#1{Eq.~(\ref{#1})}
\def\Fig#1{Fig.~\ref{#1}}

\begin{document}
\title{Observation of biradical spin coupling through hydrogen bonds}



\author{Yang He}
\affiliation{Center for Carbon-based Electronics and Key Laboratory for the Physics and Chemistry of Nanodevices, Department of Electronics, Peking University, Beijing 100871, China}

\author{Na Li}
\affiliation{Center for Carbon-based Electronics and Key Laboratory for the Physics and Chemistry of Nanodevices, Department of Electronics, Peking University, Beijing 100871, China}

\author{Ivano E. Castelli}
\affiliation{Department of Energy Conversion and Storage, Technical University of Denmark, DK-2800 Kgs. Lyngby, Denmark}

\author{Ruoning Li}
\affiliation{Center for Carbon-based Electronics and Key Laboratory for the Physics and Chemistry of Nanodevices, Department of Electronics, Peking University, Beijing 100871, China}

\author{Yajie Zhang}
\affiliation{Center for Carbon-based Electronics and Key Laboratory for the Physics and Chemistry of Nanodevices, Department of Electronics, Peking University, Beijing 100871, China}

\author{Xue Zhang}
\affiliation{Center for Carbon-based Electronics and Key Laboratory for the Physics and Chemistry of Nanodevices, Department of Electronics, Peking University, Beijing 100871, China}

\author{Chao Li}
\affiliation{Center for Carbon-based Electronics and Key Laboratory for the Physics and Chemistry of Nanodevices, Department of Electronics, Peking University, Beijing 100871, China}

\author{Bingwu Wang}
\affiliation{Beijing National Laboratory for Molecular Science, College of Chemistry and Molecular Engineering, Peking University, Beijing 100871, China}

\author{Song Gao}
\affiliation{Beijing National Laboratory for Molecular Science, College of Chemistry and Molecular Engineering, Peking University, Beijing 100871, China}

\author{Lianmao Peng}
\affiliation{Center for Carbon-based Electronics and Key Laboratory for the Physics and Chemistry of Nanodevices, Department of Electronics, Peking University, Beijing 100871, China}

\author{Shimin Hou}
\affiliation{Center for Carbon-based Electronics and Key Laboratory for the Physics and Chemistry of Nanodevices, Department of Electronics, Peking University, Beijing 100871, China}

\author{Ziyong Shen}
\affiliation{Center for Carbon-based Electronics and Key Laboratory for the Physics and Chemistry of Nanodevices, Department of Electronics, Peking University, Beijing 100871, China}

\author{Jing-Tao L\"{u}}
\email{jtlu@hust.edu.cn}
\affiliation{School of Physics, Institute for Quantum Science and Engineering, and Wuhan National High Magnetic Field Center, Huazhong University of Science and Technology, 430074 Wuhan, China}

\author{Kai Wu}
\email{kaiwu@pku.edu.cn}
\affiliation{Beijing National Laboratory for Molecular Science, College of Chemistry and Molecular Engineering, Peking University, Beijing 100871, China}

\author{Per Hedeg\aa rd}
\email{hedegard@nbi.ku.dk}
\affiliation{Niels Bohr Institute, University of Copenhagen, DK-2100 Copenhagen, Denmark}

\author{Yongfeng Wang}
\email{yongfengwang@pku.edu.cn}
\affiliation{Center for Carbon-based Electronics and Key Laboratory for the Physics and Chemistry of Nanodevices, Department of Electronics, Peking University, Beijing 100871, China}
\affiliation{Beijing Academy of Quantum Information Sciences, Beijing 100193, China}


\date{\today}

\begin{abstract}
Investigation of intermolecular electron spin interaction is of fundamental importance in both science and technology.
Here, radical pairs of all-trans retinoic acid molecules on Au(111) are created using an ultra-low temperature scanning tunneling microscope. Antiferromagnetic coupling between two radicals is identified by magnetic-field dependent spectroscopy.
The measured exchange energies are from 0.1 to 1.0 meV. The biradical spin coupling is mediated through O-H$\cdots$O hydrogen bonds, as elucidated from analysis combining density functional theory calculation and a modern version of valence bond theory.
\end{abstract}


\maketitle



Electron spin interactions play pivotal roles in both fundamental science and applied technology. Spin-related quantum effects have been intensively investigated in solid materials. Electron spins can couple via exchange, double-exchange, superexchange and conduction-electron-mediated Ruderman-Kittel-Kasuya-Yosida interactions. Scanning tunneling microscope (STM) is a powerful tool to investigate spin properties for its capability of high-resolution spectroscopy at the single-atom level \cite{LiJiutao1998,Madhavan1998}.
Spin interaction between magnetic metal atoms were intensively investigated by STM \cite{Hirjibehedin2006,Bork2011,Spinelli2015,Ternes2017,Choi2019}. Data storage and logic operation have been realized based on the manipulation of spin states of nanometals \cite{Loth2012,Ako2011}.

By probing Kondo resonance and spin excitation, magnetic properties of organic molecules were characterized using STM \cite{ZhaoAidi2005,Komeda2011,Mugarza2011,Franke2011,Minamitani2012,Zhang2013,Verlhac2019}. Intermolecular spin interactions were rarely investigated even though they are of fundamental importance \cite{Chen2008}. The hydrogen bond is an unique attractive interaction in molecular crystals and biological structures. Spin couplings through hydrogen bonds have been studied by electron spin resonance and magnetic neutron diffraction\cite{Pontillon1999}. These methods average out spectroscopic signals from a large number of molecules including their mutual interactions. The experimental investigation at the single-molecule level has not been reported.

\begin{figure}
\centering
\includegraphics[width=8cm] {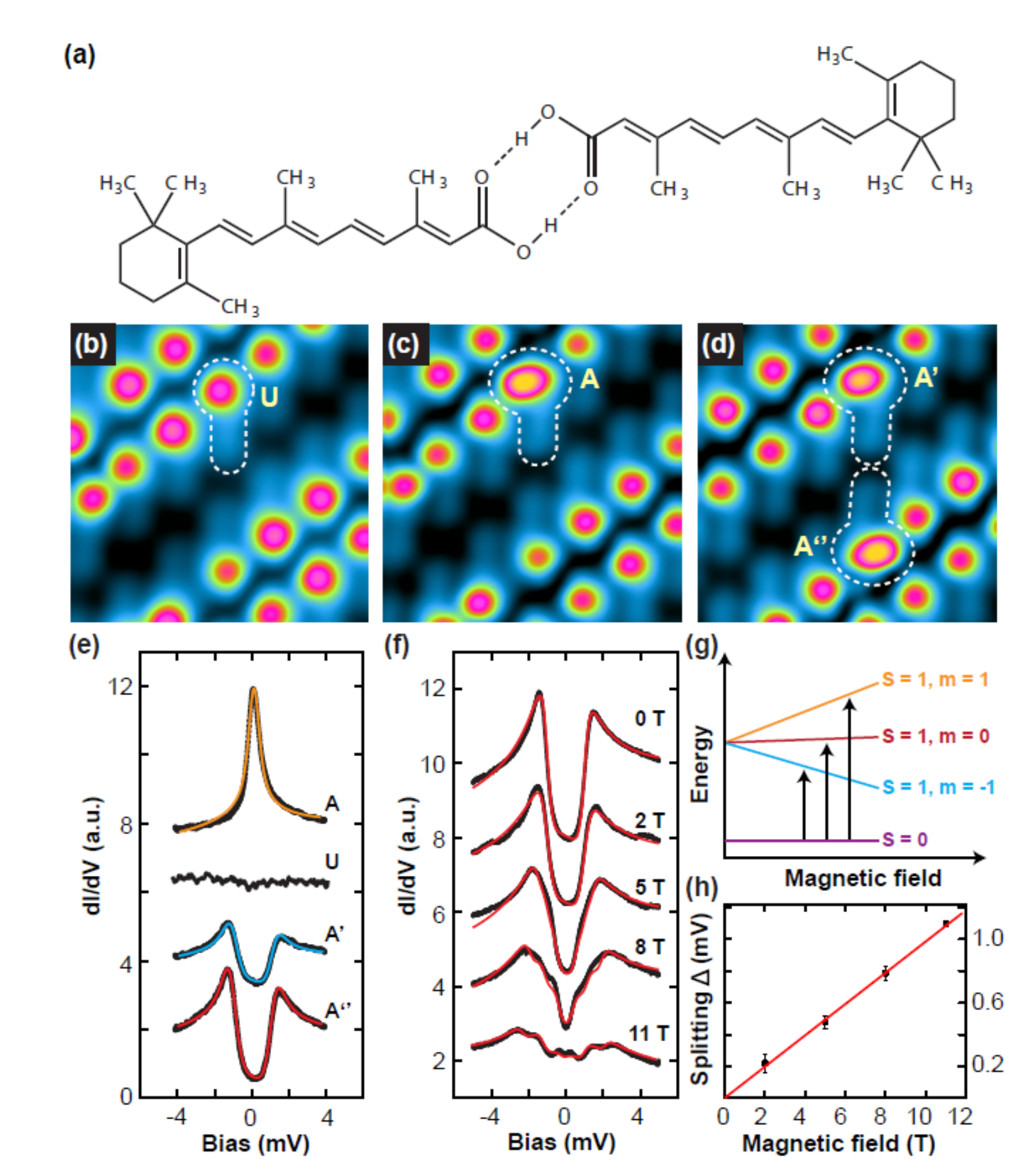}
\caption{Formation and characterization of a hydrogen-bonded ReA radical pair. (a) Chemical structure of a ReA dimer. (b-d) STM images showing the formation process of a ReA radical pair. The molecular non-magnetic, spin-carrying, and spin-coupled states are denoted by U, A, and A'A'', respectively. Imaging parameters: (b) $V = -4$ mV, $I = 16$ pA; (c) $V = -10$ mV, $I = 16$ pA; (d) $V = -10$ mV, $I = 16$ pA. The sizes of all three STM images are 4.5 $\times$ 4.5 nm$^2$. (e) $dI/dV$ spectra measured over the centers of molecular bulky heads of different states (U, A, A', A''). The STM is operated at $V = 11$ mV and $I = 0.5$ nA before opening the current feedback loop when taking constant-height spectroscopic measurements. (f) Magnetic-field dependent $dI/dV$ spectra recorded on the head group of state  A'' with the set point of -5.7 mV and 1.2 nA. In both (e) and (f), all black dotted curves were experimentally obtained and solid color curves represent fits of the spectra. All plots are vertically shifted for clarity. (g) Illustration of spin transitions between singlet and triplet states in magnetic fields.
(h) Extracted Zeeman splitting as a function of magnetic field. A linear fit yields the $g$ = 1.7.}
\label{Characterization2}
\end{figure}

All-trans retinoic acid (ReA, Fig.~\ref{Characterization2}a) can be transited into a radical via biphotonic processes \cite{Li2013}. Here, we use a low-temperature STM to construct radical pairs of ReA molecules on Au(111). The spin generation, spin distribution and exchange interaction of radical pairs are studied by a combination of STM and density functional theory (DFT) calculations. The singlet ground state of the hydrogen-bonded radical pair is determined and the exchange energy ranges from 0.1 to 1.0 meV. Analysis based on a modern version of valence bond theory (VBT) reveals that the biradical superexchange antiferromagnetic (AFM) coupling is mediated by hydrogen bonds.

The experiments are performed using an ultrahigh vacuum scanning tunneling microscope (UNISOKU USM-1300-$^3$He) with magnetic field up to 11 T. Single crystalline Au(111) (MaTecK GmbH) surfaces are cleaned by cycles of Argon ion sputtering and annealing. ReA molecules (Sigma Aldrich) are evaporated from a homemade tantalum boat onto Au(111) kept at room temperature. The measurements are conducted at 0.5 to 0.6 K with an etched tungsten tip. Differential conductance ($dI/dV$) spectra are collected by a lock-in amplifier with the modulation voltage of 0.06 mV. All STM images are slightly processed using the software of WSxM \cite{WSxm}.

ReA molecules interact with each other mainly through O-H$\cdots$O hydrogen bonds, as shown by the optimized model of a ReA dimer (Fig.~\ref{Characterization2}a). At high coverage, ReA molecules arrange themselves in parallel dimers (Fig.~\ref{Characterization2}b) \cite{Karan2013}. The molecular bulky 1,3,3-trimethylcyclohexene group appears as the highest protrusion, denoted as head group. The rest of the molecule, 3,7-dimethylnona-2,4,6,8-tetranoic acid, appears lower and is denoted as polyene. ReA molecules could be switched to spin-carrying states by positioning the tip over the molecular bulky heads and shifting the sample voltage from -2.0 to -2.8 V with current feedback disabled \cite{Karan2016}. After switching the ReA molecule (marked by U) to a spin-carrying state (marked by A), the head group appears as an elliptical protrusion and is around 50 pm higher than others (Fig.~\ref{Characterization2}c). Same to the previous work \cite{Karan2016}, the molecule can be further switched to two other magnetic states, which are presented in Fig. S1 in the supplemental material (SM). $DI/dV$ spectra measured over centers of the molecular bulky heads of U and A states are presented in Fig.~\ref{Characterization2}e. The spectrum of U state is featureless. Meanwhile, the spectrum of the switched A state shows a zero bias resonance. We attribute the resonance to Kondo effect resulting from screening of the molecular spin by the conducting electrons of Au(111) \cite{Ternes2008}. The spectrum is fitted by Frota function and the fitted Kondo temperature is 4.7 K (Fig. S2 in SM).  Interestingly, a molecule in state A further changes to state A' when its partner molecule in a dimer is switched to the A'' state by the same procedure (Fig.~\ref{Characterization2}d). Correspondingly, a small voltage gap is observed in the $dI/dV$ spectrum (Fig.~\ref{Characterization2}e). A similar spectrum is obtained on the paired A''-state molecule (Fig.~\ref{Characterization2}e).

The gapped spectra originate from the singlet-triplet spin transition, which can be well interpreted by the two-impurity Anderson model. There are two electron spins in one ReA radical pair. A competition arises between screening of molecular spin by conducting electrons and inter-molecular spin coupling. The former process favors the formation of a Kondo state, while the latter prefers to form a singlet or triplet state via AFM or ferromagnetic (FM) coupling. The evolution of $dI/dV$ spectra in magnetic fields confirm that two electron spins form a local singlet in the radical pair. Fig.~\ref{Characterization2}f shows a series of spectra recorded on the head group of ReA in state A'' with the field strength up to 11 T. When increasing the strength of magnetic field, the zero-bias gap gradually splits and a clear three-fold splitting is observed when the field reaches 11 T. The energy gaps in the spectra correspond to spin transitions from the ground singlet state ($S = 0$; $m = 0$) to triplet excited states ($S = 1$; $m = 1, 0, -1$), as schematically shown in Fig.~\ref{Characterization2}g. All spectra of spin transitions are fitted using the scattering theory proposed by Appelbaum $et$ $al.$, revised and implemented by Ternes \cite{Ternes2015}. Since there is still a strong spin-flip Kondo scattering at the spin excitation step at the low experimental temperature, we have replaced the logarithmic function by the Frota function in the fitting, which provides a
better fitting of the spectra at the two steps \cite{Bergmann2015}. The Hamiltonian is
\begin{equation}
H = \sum_{i=A',A''} g \mu_B \vec{B}\cdot \vec{S}^i + {J}_{} \vec{S}^{A'}\cdot \vec{S}^{A''}.
\end{equation}
Here, the index $i$ represents molecules in the dimer, $J$ is the Heisenberg exchange coupling, $g$ is the Land\'e g-factor and $\mu_B$ is the Bohr magneton.
For $B=0$, we get $J_{}=1$ meV from the fitting. An average effective $g$-factor $g=1.7$ is extracted by fitting the magnetic-field dependent spectra (Fig.~\ref{Characterization2}h). All other fitting parameters are listed in Fig. S3 in SM.

\begin{figure}
\centering
\includegraphics[width=8cm] {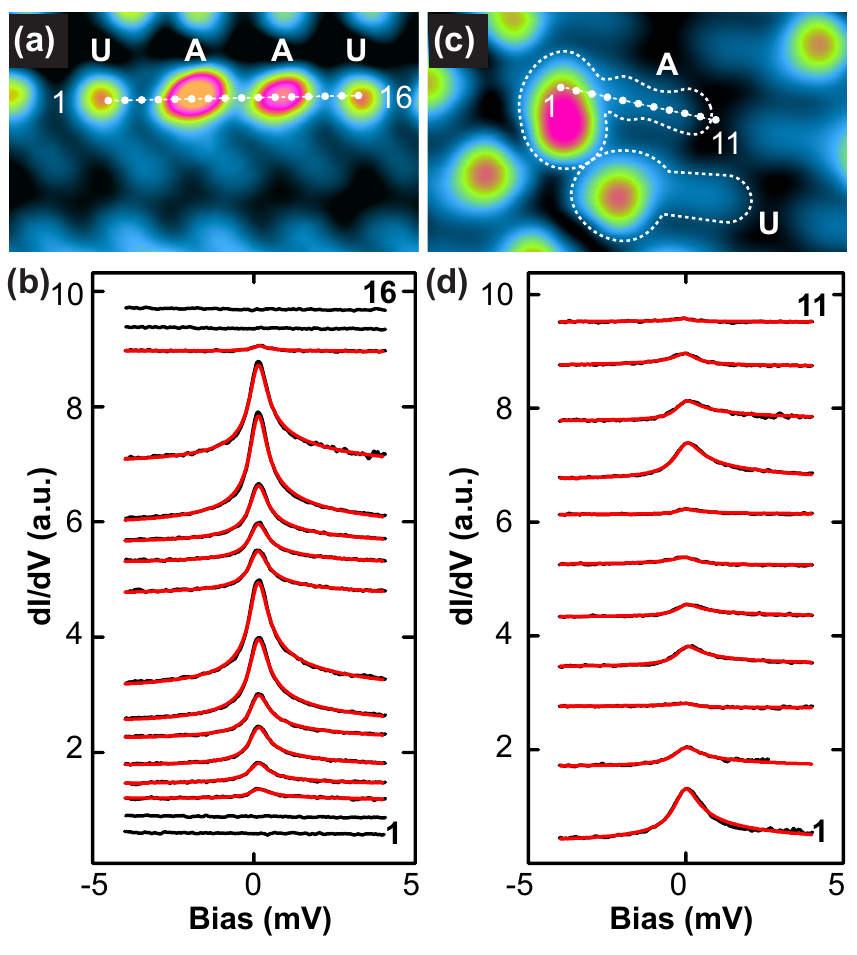}
\caption{{\label{LDOS} Experimental confirmation of spin coupling through hydrogen bonds.} (a) STM image comprising two A-state ReA molecules packed side by side with their molecular bulky heads higher than others (-11 mV, 0.10 nA; 4.1 $\times$ 2.4 nm$^2$). (b) $DI/dV$ spectra taken over the sixteen white-dot positions in (a). (c) STM image of A-state ReA molecule.
  (-11 mV, 0.10 nA; 2.3 $\times$ 1.5 nm$^2$). (d) $DI/dV$ spectra measured over the eleven white-dot positions along the line in (c). For all spectra, the STM is operated at -11 mV and 0.53 nA before opening the feedback loop. }
\label{LDOS}
\end{figure}

When two ReA radicals are not connected through hydrogen bonds, only the Kondo resonance of single ReA radical is observed. One example is shown in Fig.~\ref{LDOS}a, where two ReA molecules packed side by side are manipulated to magnetic A states. The measured spectra at different positions have not shown any feature of spin coupling (Fig.~\ref{LDOS}b). The fitted Kondo temperatures of the spectra are presented in Fig. S4 in SM. The $dI/dV$ measurements at eleven white-dotted positions marked in Fig.~\ref{LDOS}c indicate that the electron spin is delocalized on the ReA radical (Fig.~\ref{LDOS}d). The spin delocalization has been reported for other organic molecules \cite{Patera2019,Zheng2020,Mishra2020,Wang2021}. The fitted asymmetric factors and Kondo temperatures are listed in Fig. S5 in SM. The spin delocalization is found for all ReA radicals and the related measurements for two other single ReA radicals and one radical dimer are presented in Figs. S6 and S7 in SM. The spin delocalization allows spin coupling through hydrogen bonds. In the following, we perform DFT calculations (SM) and VBT analysis to elucidate the mechanism of spin delocalization, position-dependent Kondo spectra of sing ReA radical and spin coupling of a ReA radical pair through hydrogen bonds.
\begin{figure}
\centering
\includegraphics[width=8cm] {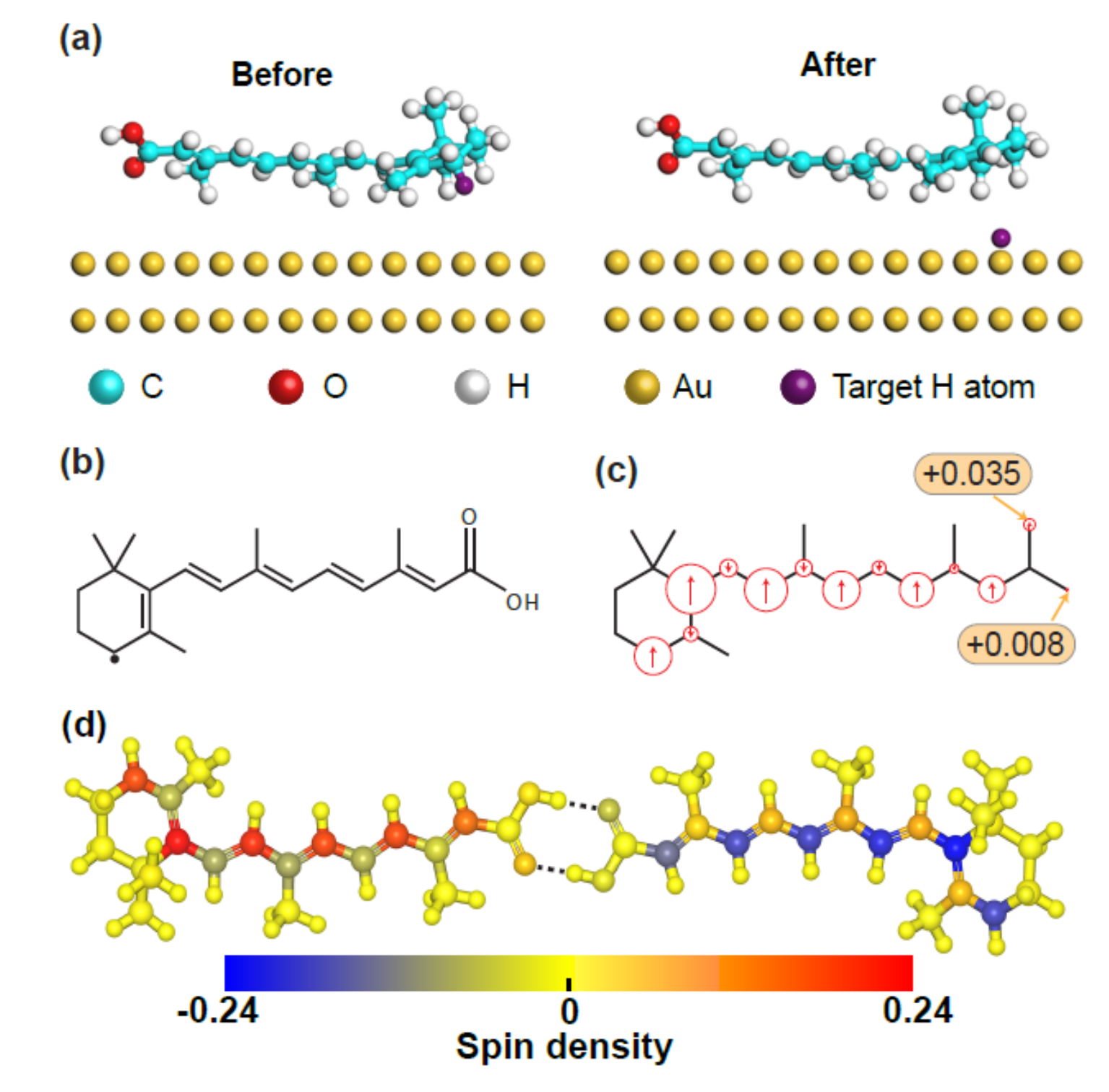}
\caption{Mechanism of biradical spin coupling through hydrogen bonds. (a)  Optimized structures of a ReA molecule on Au(111) and that after removing the purple H atom.
(b) A model of ReA radical formed by removing one H atom at dotted position. (c) Calculated spin distribution of the ReA radical on Au(111).  The spin density is represented by red arrows with different length. (d) Optimized structure of a ReA radical pair.}
\label{Maps}
\end{figure}

\begin{figure*}
\centering
\includegraphics[width=14cm] {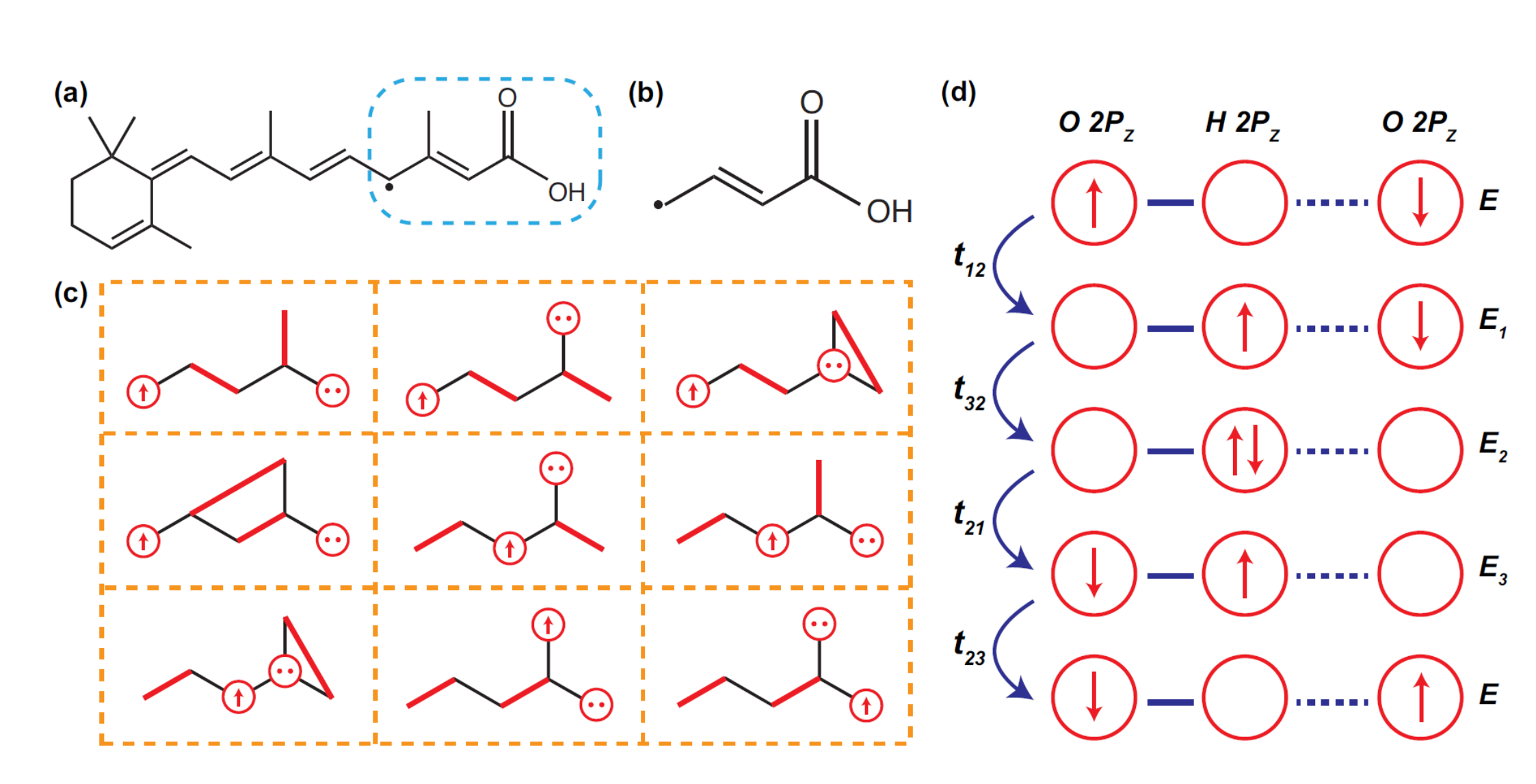}
\caption{ VBT analysis of biradical spin coupling through hydrogen bonds. (a) One resonant structure of the ReA radical shown in Fig.~ 3a. (b) Simplified molecular model of the ReA radical for VBT analysis. (c) Nine valence bond states of (b), which reveals that the spin polarity of oxygen atoms is same to that of the ReA radical. (d) One of the six exchange processes, where two spins on oxygen $2p_z$ orbitals on separate molecules are being exchanged. The amplitudes of the hops are denoted by $t_{ij}$, and the intermediate energies are $E_i$.}
\label{BasicExchange}
\end{figure*}

 The dehydrogenation process has been used to generate spin on the graphene nanostructure \cite{Li2019}. In a previous report, a magnetic ReA molecule was considered as a radical which was generated by H-atom transfer and sigmatropic reaction \cite{Bocquet2019}. However, the mechanism leads to a localized spin. Here, we propose another possible activation mechanism which involves only the transfer of H atom from ReA to Au(111) (Figs. 3a). A neutral radial is formed after losing the hydrogen atom at the radical carbon position, as shown in Fig.~\ref{Maps}b. The calculated activation barrier is roughly 3 eV (Fig. S8 in SM). A delocalized spin distribution is found on the ReA radical (Fig. S9 in SM). The calculated spin density at each atom of a ReA radical on Au(111) is represented with red arrows of different length in Fig.~\ref{Maps}c. The values are listed in Table 1 in SM. From molecular head to the carboxyl group, the spin densities of six carbon atoms marked by red spin-up arrows (Fig. 3c) are 0.19, 0.24, 0.21, 0.18, 0.17 and 0.13, which shows a clear decay. After removing the H atom, the reactive C atom interacts more strongly with the substrates than other atoms. For these two reasons, the amplitude of Kondo resonance decays from molecular head to the carboxyl group. The oscillatory behavior of the Kondo peak originates from that the amplitude of spectra at methyl groups is stronger than that of other positions, which might be due to structure dependent molecule-substrate coupling. The asymmetric factor $\phi$ is determined by the ratio between the electron tunneling into the Kondo resonance and substrate. So, position-dependent spin density and coupling between ReA and Au(111) are tentatively suggested to interpret why the fitted asymmetry factor and Kondo temperature (Fig. S3 in SM) are not constant. Two ReA radicals form a singlet ground state via AFM interaction (Fig.~\ref{Maps}d). The calculated exchange energy of the optimized dimer is 0.50 meV, which nicely matches the experimental result ranging from 0.1 to 1.0 meV (Fig. S11 in SM). 


Furthermore, the analysis based on a modern version of VBT \cite{RN428,RN427,RN426,Pedersen2013} reveals the physical nature of biradical spin coupling. A brief introduction about the theory is shown in the section of valence bond theory in SM (Figs. S12 to S15). The ReA radical is conjugated and one of its resonant formulas is shown in Fig.~\ref{BasicExchange}a. Only the marked part of the ReA radical is used in the VBT model (Fig.~\ref{BasicExchange}b), which simplifies the theoretical analysis. A total of nine valence bond states are displayed in Fig.~\ref{BasicExchange}c. A red line represents a singlet electron pair residing on the atoms at the ends of the line. The electron spin can move to two oxygen atoms in the -COOH group and their spin polarities are the same as that of the molecule. It is less probable for the electron spin to reach the hydroxyl oxygen atom because the process involves breaking the lone pair of that oxygen. This well explains the DFT calculated results that the spin density of the carbonyl oxygen (+0.035) is much higher than that of the hydroxyl oxygen (+0.008). The exchange interaction between these two types of oxygen atoms in an O-H$\cdots$O hydrogen bond determines spin coupling between two ReA radicals. There are in total six super-exchange processes and one of them is illustrated in Fig.~\ref{BasicExchange}d. Two electrons switch orbitals in four hops via intermediate occupation of the $2p_z$ orbital of a hydrogen atom. From these exchange processes, we can conclude that two electron spins located at the oxygen atoms should be opposite. Since the spin polarities of each oxygen atom and the molecule are identical, two ReA radicals form a singlet ground state via the AFM super-exchange interaction.

The order of exchange energy can be estimated from the VBT. The biradical spin coupling acts as a perturbation to the uncoupled molecules. The theory includes hopping terms between $p_z$ orbitals and Coulomb interaction between charged atoms. The effective Hamiltonian coupling the $\pi$-system in the O-H$\cdots$O hydrogen bond (Fig.~\ref{BasicExchange}d) is
\begin{equation}
\label{Heff}
  H_{\rm exch} = -J_H \left(\frac{1}{4}-\vec S_1\cdot\vec S_2
  \right).
\end{equation}
Here the exchange coupling parameter $J_H$ is given by
\begin{align}\label{ExConst}
J_H &= \frac{12t_{1}^2 t_{2}^2}{(E_1-E)(E_2-E)(E_3-E)},
\end{align}
where $E_i$ $(i=1,2,3)$ is the energy of the intermediate state and $E$ is the energy of the unperturbed ground state. The constant is positive resulting again in an AFM exchange coupling across the hydrogen bond. It is exponential in the distance between atoms, which is the reason why an exponentially decaying exchange coupling is found in the DFT result (Fig. S10 in SM).
We use $\vec S_1$ ($\vec S_2$) and $\vec S_a$ ($\vec S_b$) to represent the spin operators of the oxygen atom and whole ReA radical, respectively. The Wigner-Eckard theorem ensures that these two operators are proportional. The proportionality constant can be determined by evaluating their expectation value in the ground state of the molecules. We get $\vec S_1 = M_1 \vec S_a
$, with $M_1 = 2 \langle S_1^z\rangle$. The coupling between two ReA radicals through hydrogen bonds thus becomes
\begin{equation}\label{HeffS}
  H_{\rm exch}  = - J_H M_1 M_2 \left(\frac{1}{4}-\vec S_a\cdot\vec S_b\right).
\end{equation}
In vacuum, the spin densities of two oxygen atoms are 0.056 and 0.013 (Table 1 in SM), respectively. An extra factor 2 is introduced because there are two hydrogen bonds in a dimer. Therefore, the singlet-triplet splitting energy is around $2J_H\times 2 \times 0.056 \times 2 \times 0.013 = 0.0058J_H$. When comparing to DFT calculated exchange energy of 0.5 meV, we estimate the value of $J_H$ is roughly 0.1 eV. When considering the molecule-surface interaction the spin densities of carbonyl and hydroxy oxygen atoms reduce to 0.035 and 0.008, which leads to an exchange energy reduction from 0.5 meV in vacuum to 0.22 meV on Au(111).

In summary, biradical spin coupling through hydrogen bonds has been studied by a combination of LT STM, DFT calculations and VBT analysis. The AFM interaction between two ReA radicals is identified by magnetic-field dependent spectroscopy. The measured exchange energies range from 0.1 to 1.0 meV, which are in agreement with the DFT calculated results. The VBT analysis reveals that spin coupling in ReA radical pairs is through O-H$\cdots$O hydrogen bonds via super-exchange processes. This work provides a way to study spin properties of single radical pairs.

\vspace{0.3cm}

\begin{acknowledgments}
This work was supported by the Ministry of Science and Technology (Nos. 2017YFA0205003, 2018YFA0306003, 2017YFA0204702, 2017YFA0403501) and National Natural Science Foundation of China (Nos. 21433011, 91527303, 21522301, 61621061, and 21873033).
\end{acknowledgments}
\vspace{0.3cm}
Y. He, N. Li, I. E. Castelli and R. N. Li contributed equally to this work.

\appendix
\def\Eq#1{Eq.~(\ref{#1})}
\def\Fig#1{Fig.~\ref{#1}}
\setcounter{figure}{0}

\newcommand*\mycommand[1]{\texttt{\emph{#1}}}
\renewcommand{\thefigure}{S\arabic{figure}}

\section*{Supplemental Materials}
\begin{figure}
	\centering
	\includegraphics[width=6cm] {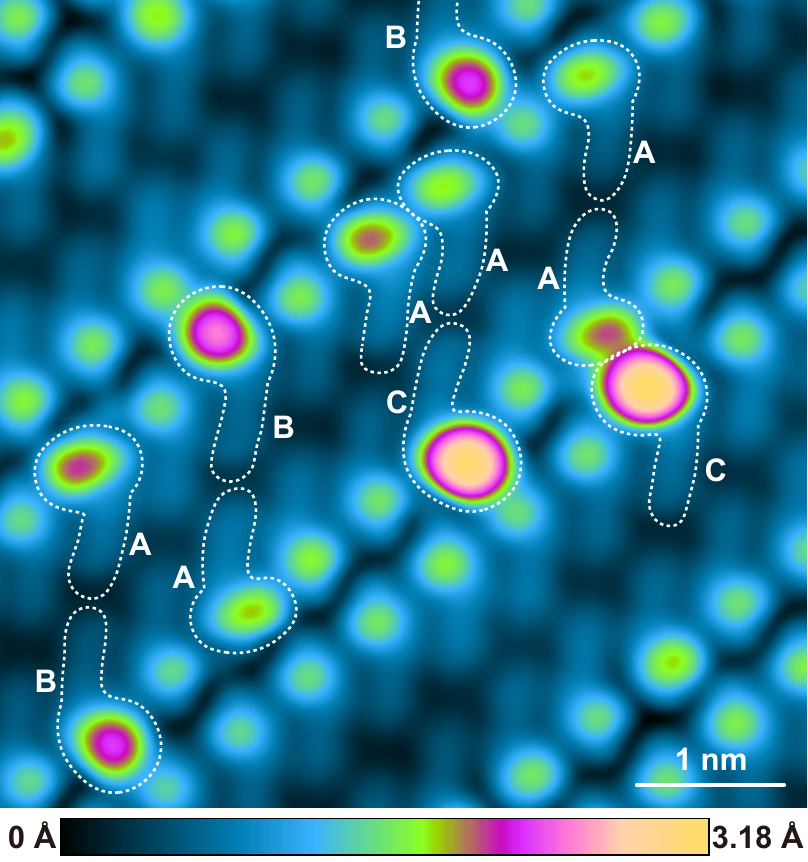}
	\caption{STM image of ReA molecule and its three manipulated magnetic states which are labeled and marked by white dotted lines ($V$ = -10 mV, $I$ = 10 pA). They can be differentiated by STM image and $dI/dV$ spectrum. For state A, the molecular head group appears as an elliptical protrusion. Rounded shapes are viewed in state B and C at molecular bulky heads. State C is highest in STM images. State A has relatively low Kondo temperature ranging from around 3 to 10 K. For example, the ReA radicals shown in Fig. 1e, Figs. 2a and b are all at state A and their Kondo temperatures are 4.2, 8.9 and 6.2 K, respectively. Because of the long-range reconstruction, the coupling between ReA radical and substrate varies at different adsorbed position, which might lead to the different Kondo temperature for the same-state of ReA radical. Same to Ref. 26, the Kondo temperature for state B is similar to that of state A. For state C, the Kondo temperature is much higher and above 20 K.
	}
	\label{ReAVB}
\end{figure}

\begin{figure}
	\centering
	\includegraphics[width=0.45\textwidth] {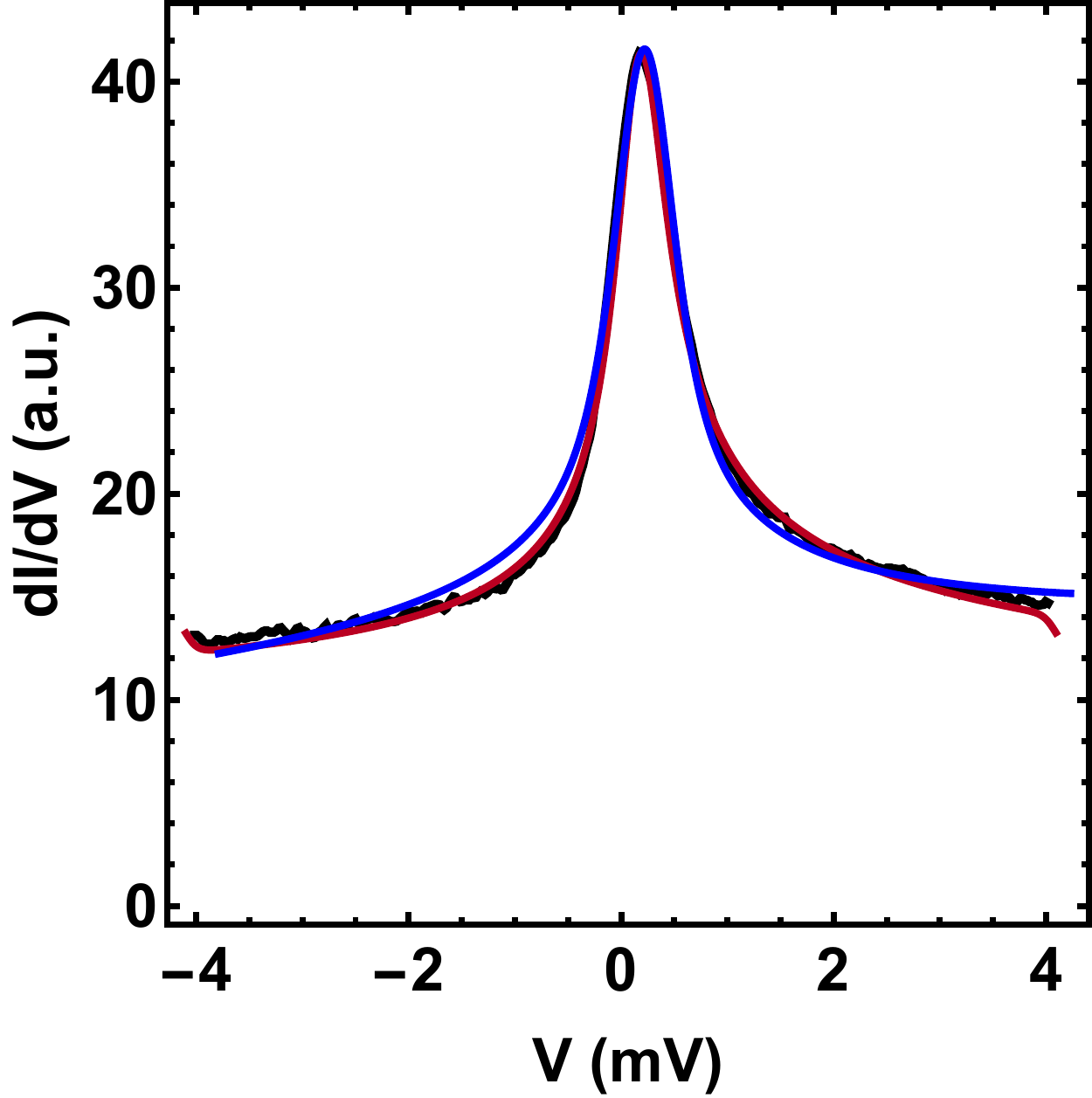}
	\caption{
		Fitting of the Kondo spectrum using different methods. The black line is experimental data, the red line is fitting using the Frota function following Gruber $et$ $al.$\cite{Gruber2018}, while the blue line is fitting using the logarithmic function\cite{Appelbaum1967,Anderson1966,Ternes2015}.}
	\label{fig:S4}
\end{figure}

\section{Fitting of Kondo spectra}
We fit the Kondo spectrum (Fig. 1(e)A) using the method described in Ref.~\cite{Gruber2018}. Specifically, we use the Frota function\cite{Gruber2018,Frota1986}
\begin{equation*}
	{\rm Frota}(V) \propto {\rm Im}\left[i e^{i\phi}\sqrt{\frac{i\Gamma_{\rm Frota}}{eV-E_K+i\Gamma_{\rm Frota}}}\right],
\end{equation*}
with a lock-in $V_m=0.06$ mV and a temperature broadening $T=610$ mK.  To include these broadening, we make convolution of the Frota function with both a lock-in modulation:
\begin{equation*}
	\xi_{\rm lock-in}(V) = \left\{
	\begin{array}{cc}
		\frac{2\sqrt{V_m^2-V^2}}{\pi V_m^2}, & (|V|\le V_m)\\
		0, & {\rm otherwise}
	\end{array}
	\right.
\end{equation*}
and the temperature broadening
\begin{equation*}
	\xi_T(V) = -\frac{\partial f(V)}{\partial T},
\end{equation*}
with
\begin{equation*}
	f(V) = \frac{1}{e^{V/k_BT}+1}
\end{equation*}
the Fermi-Dirac distribution function, $k_B$ the Boltzmann constant.
The following parameters are obtained from the fitting: $\phi=-0.2$, $\Gamma_{\rm Frota}=0.16$, with a linear background slope of $b=-0.05$. We obtain a Kondo temperature
\begin{equation}
	T_K=\frac{2.542 \Gamma_{\rm Frota}}{k_B}= 4.7\ {\rm K}
\end{equation}
from the fitting. 
For comparison, we have also performed fitting using the perturbation approach of Appelbaum and Anderson\cite{Appelbaum1967,Anderson1966,Ternes2015}.
We use the coefficient of determination
\begin{align*} 
R^2 = 1-\frac{\sum_i (y_i-f_i)^2}{\sum_i (y_i-\bar{y})^2}
\end{align*}
to characterize the quality of the fitting.
As pointed out in Ref.~\cite{Frota1986}, the Frota function gives a better fitting than the perturbation approach. Here, their $R^2$ are 0.995 and 0.987, respectively.
The same procedure (using Frota function) is applied to fit the position dependent Kondo spectra in Figs. 2(b) and (d). The resulting parameters are listed in Figs. S4 and S5.

\begin{figure}[H]
	\centering
	\includegraphics[scale=0.55] {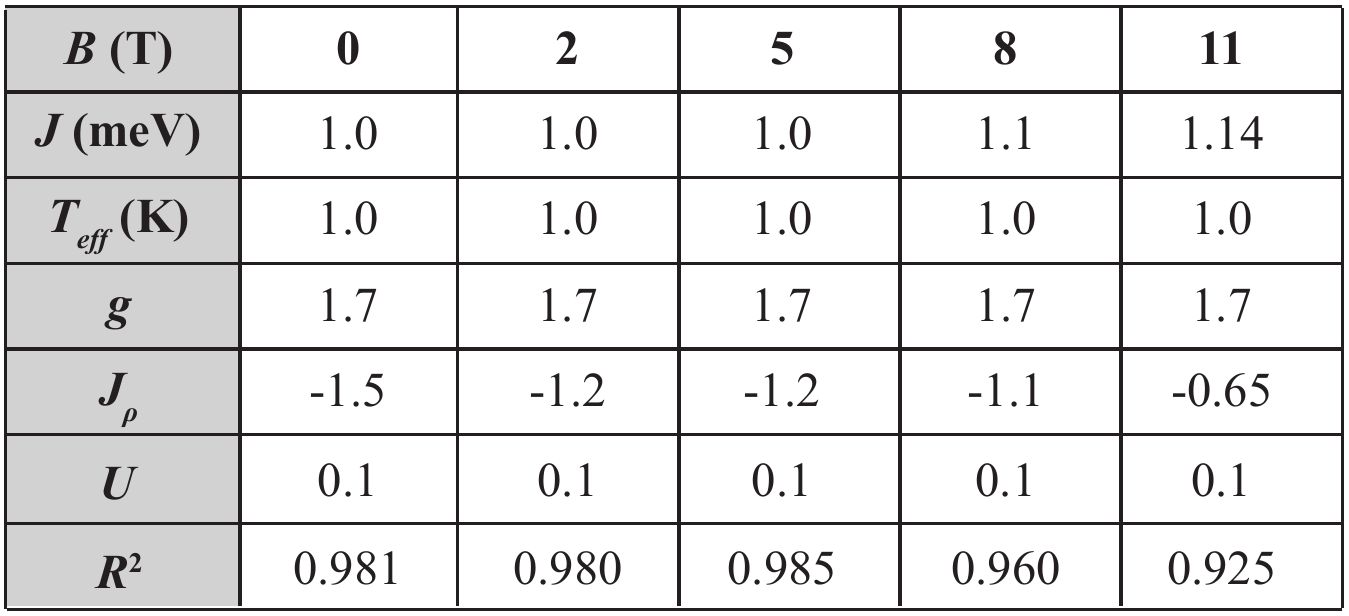}
	\caption{Fitting parameters of the field-dependent $dI/dV$ spectra shown in Fig. 1f of the main text. Here, $J_\rho$ is the dimensionless Kondo scattering parameter for the spin system with the substrate, and $U$ is the dimensionless Coulomb scattering parameter of the spin.}
	\label{fig:S7}
\end{figure}

\begin{figure}[H]
	\centering
	\includegraphics[scale=0.48] {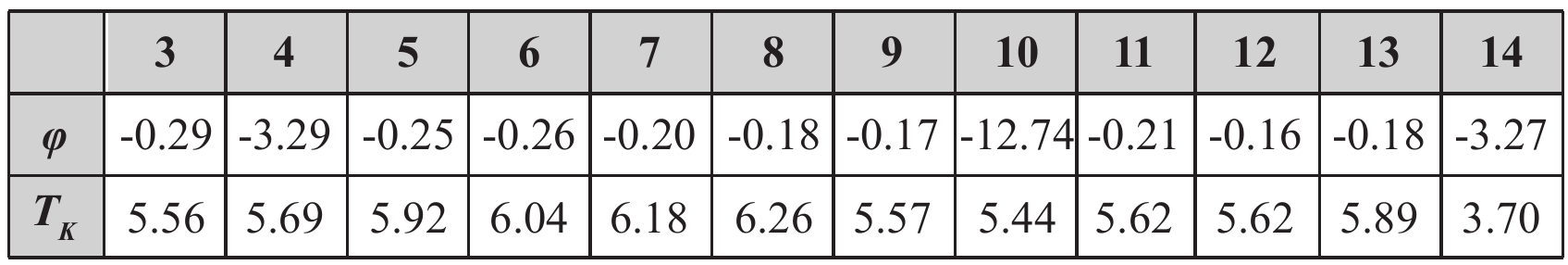}
	\caption{Fitting parameters of the position-dependent Kondo spectra shown in Fig. 2b of the main text.}
	\label{fig:S7}
\end{figure}

\begin{figure}[H]
	\centering
	\includegraphics[scale=0.5] {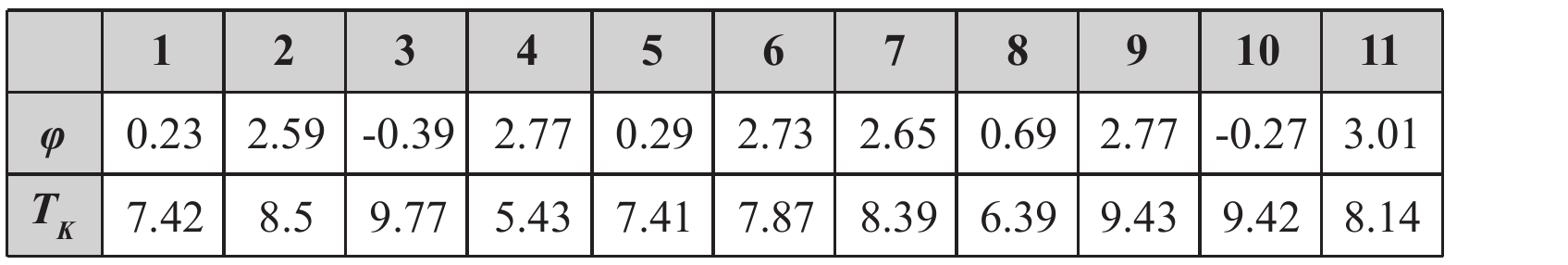}
	\caption{Fitting parameters of the position-dependent Kondo spectra shown in Fig. 2d of the main text.}
	\label{fig:S7}
\end{figure}

\begin{figure}[H]
	\centering
	\includegraphics[width=0.5\textwidth] {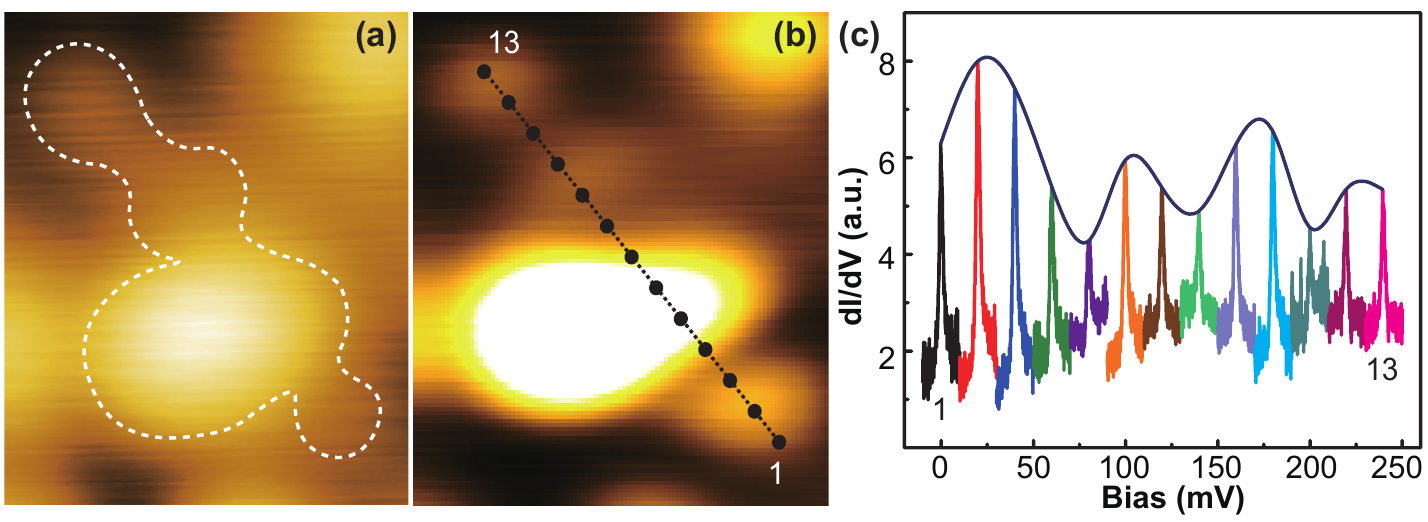}
	\caption{(a) STM image of a ReA radical (1V, 10 pA). (b) Constant-height $dI/dV$ map of the ReA radical at zero bias. (c) Position-dependent $dI/dV$ spectra, which are shifted 20 mV sequentially along the $x$-axis. The data are measured by another STM (Unisoku-1500) at 2.8 K.}
\end{figure}

\begin{figure}[H]
	\centering
	\includegraphics[scale=1] {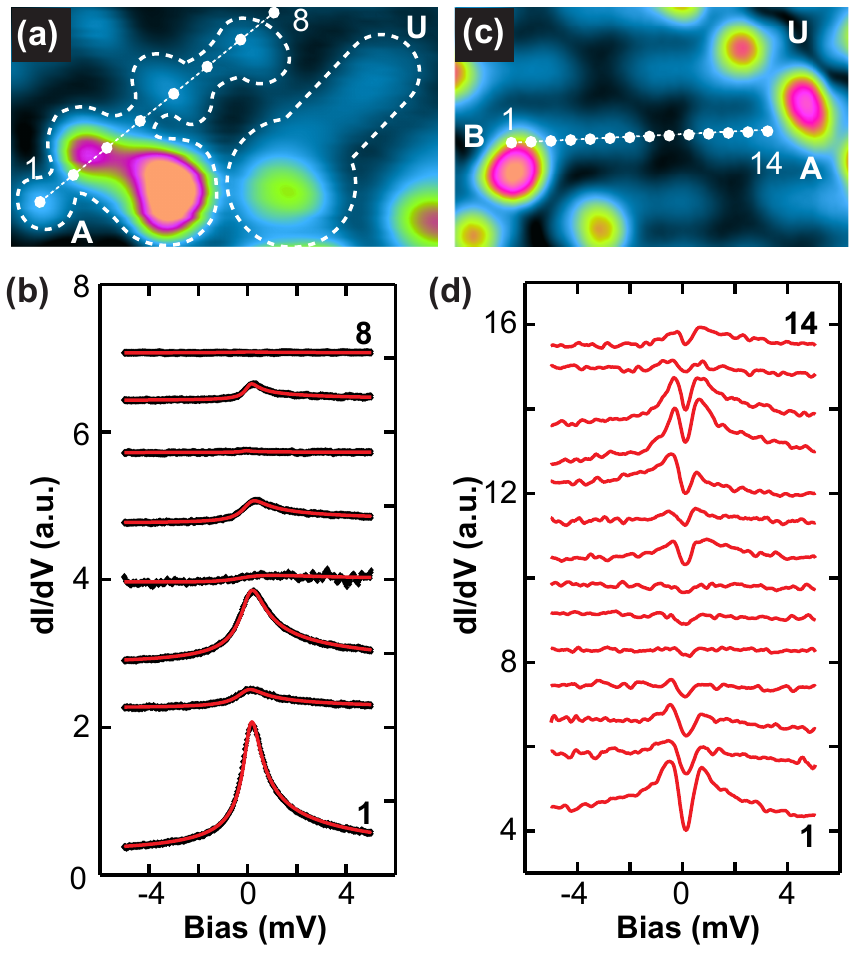}
	\caption{(a) Constant-height $dI/dV$ map for ReA  molecules in states A and U (2.3$\times$1.5 nm$^2$). (b) $dI/dV$ spectra measured over the eight white-dot positions along the line in (a). (c) STM image containing a AB-state radical dimer overlaid by white dotted line ( -11 mV, 0.1 nA; 4.1$\times$1.9 nm$^2$). (d) $dI/dV$ spectra recorded at the white-dot positions in (c).}
\end{figure}

\section{Density Functional Theory}
We have performed spin-polarized density functional theory (DFT)
calculations using the GPAW code \cite{GPAW1,GPAW2} together with the Atomistic Simulation Environment (ASE) package to build the structures \cite{HjorthLarsen2017}. The ReA molecule has been relaxed on top of a $8\times4\times4$ Au(111) supercell using BEEF-VdW as exchange-correlation functional \cite{PhysRevB.85.235149} to include Van der Waals interactions with a Monkhorst-Pack $k$-point grid of $2\times3\times1$ and $0.2$ grid spacing. We have included a vacuum of $15$ \AA~to separate the periodic images in the direction normal to the surface and we apply a dipole correction to
separate the contribution to the electrostatic interaction between the periodic images. The periodic images in the $xy$-plane are instead separated by at least $6$ \AA. To study the coupling mechanism between 2 ReA molecules, we have fully relaxed the two molecules in vacuum using only the $\Gamma$-point.

\begin{figure}[H]
	\centering
	\includegraphics[width=8cm] {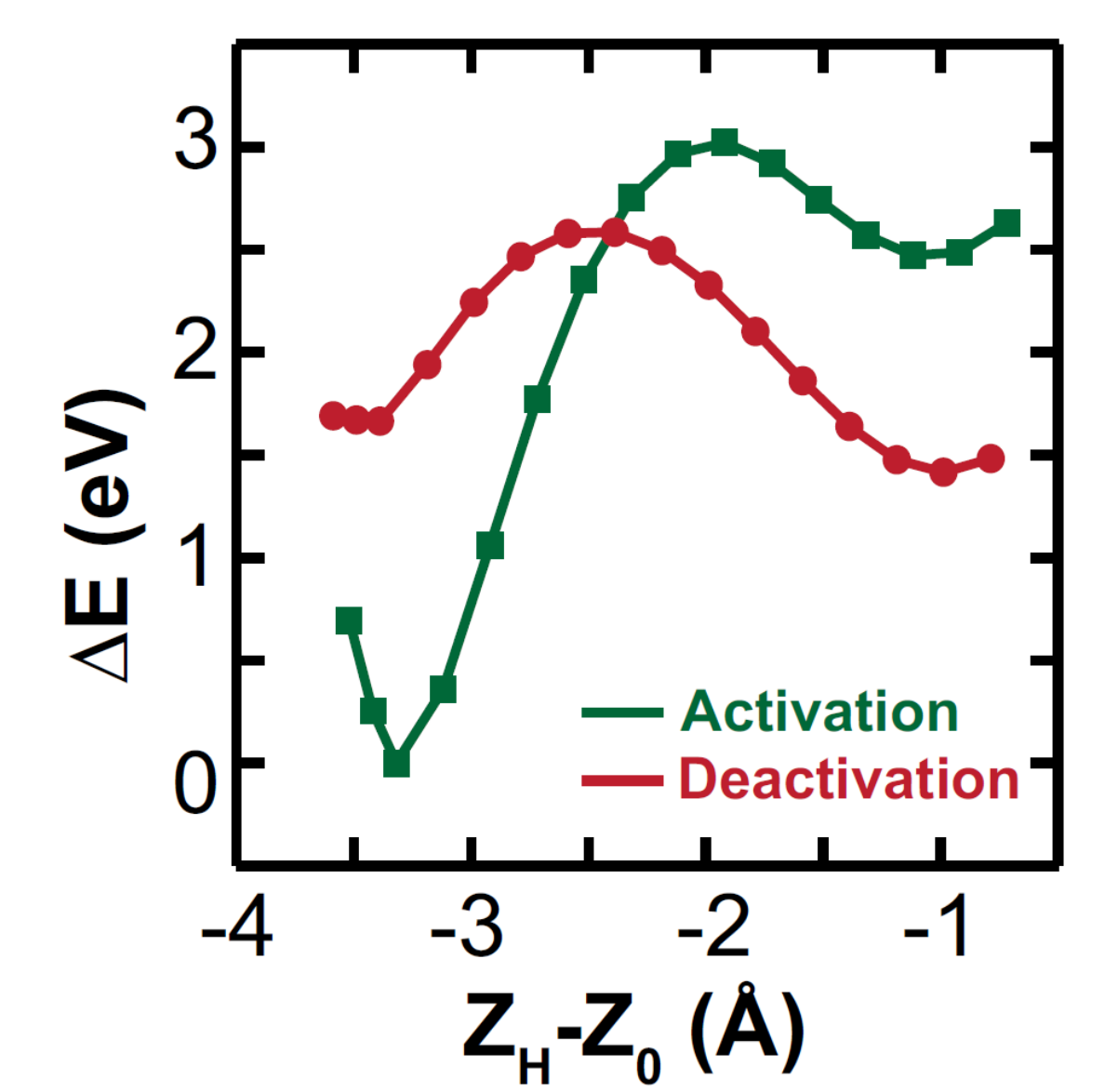}
	\caption{
		Total energy of a ReA molecule on Au(111) surface
		as a function of Hydrogen atom position as it moves from
		ReA to Au(111). In green is shown the barrier as the
		Hydrogen is moving from molecule to surface. In red is shown the barrier as
		the Hydrogen is moving from surface to molecule.}
	\label{FerroCoupling}
\end{figure}

\begin{figure}[H]
	\centering
	\includegraphics[width=0.5\textwidth] {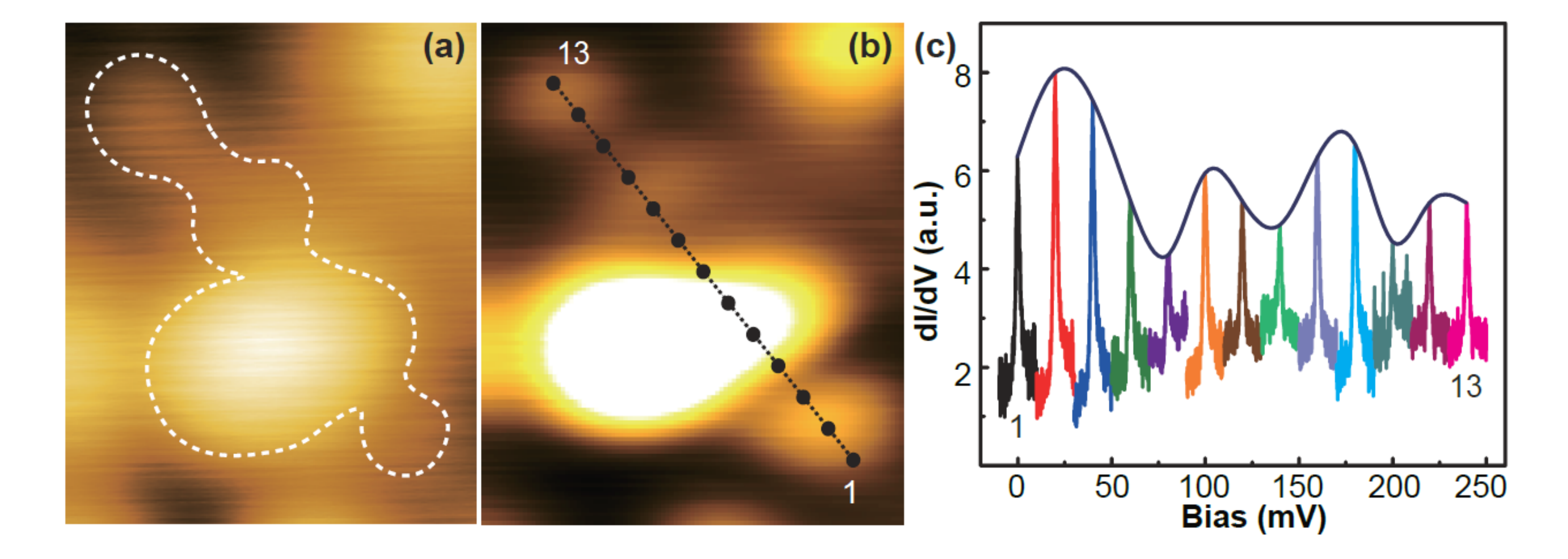}
	\caption{Calculated SOMO of a ReA radical.}
	\label{FerroCoupling}
\end{figure}

\begin{figure}
	\centering
	\includegraphics[width=8cm] {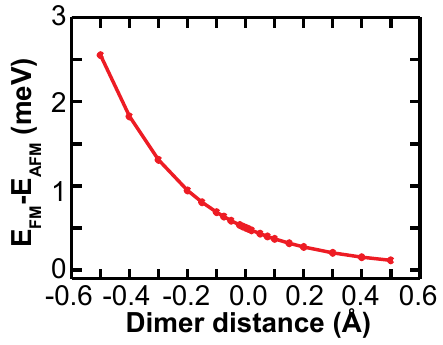}
	\caption{Calculated exchange energies of hydrogen bonds with different  O-H$\cdots$O bond length.}
	\label{FerroCoupling}
\end{figure}

\section{Exchange energies of different ReA-radical pairs}
The measured exchange energies are different for various radical pairs. The energies of two AA-state ReA dimers are 1.0 (Fig. 1f) and 0.11 meV (Fig. S11), respectively. We attribute this difference to the variation of the strength of hydrogen bonds. At different locations, the molecule-substrate interactions are different because of the long-range reconstructed substrate, which leads to the strength change of hydrogen bonds. To investigate the influence of strength of hydrogen bonds on exchange energy, we change the hydrogen-bond distance from -0.5 to 0.5 angstrom (relative to the equilibrium distance) step by step and calculated their exchange energy. The calculated results range from 0.11 to 2.55 meV (Fig. S10). The exchange energy becomes larger when hydrogen bonds get stronger. This might interpret various experimentally observed exchange energies for different dimers at the same magnetic state. The exchange energies of two AB- and one BB-state ReA radical pairs are 0.15, 0.35 and 0.29 meV. For state B, the Kondo temperature is similar to that of state A. Therefore, we have not observed significant variation of exchange energies when one or two state-A molecules are switched to state-B molecules.
When there is a state-C ReA in a radical pair (BC, AC or CC), the spin exchange is not experimentally obtained. For example, $dI/dV$ spectra on molecules of a BC pair are presented in Fig. S11. Only Kondo resonances are observed. The coupling between state C and Au(111) is strong enough to screen molecular electron spin efficiently. Therefore, spin coupling through hydrogen bonds between two molecules get weaker and is not observed.

\begin{figure}
	\centering
	\includegraphics[width=0.45\textwidth] {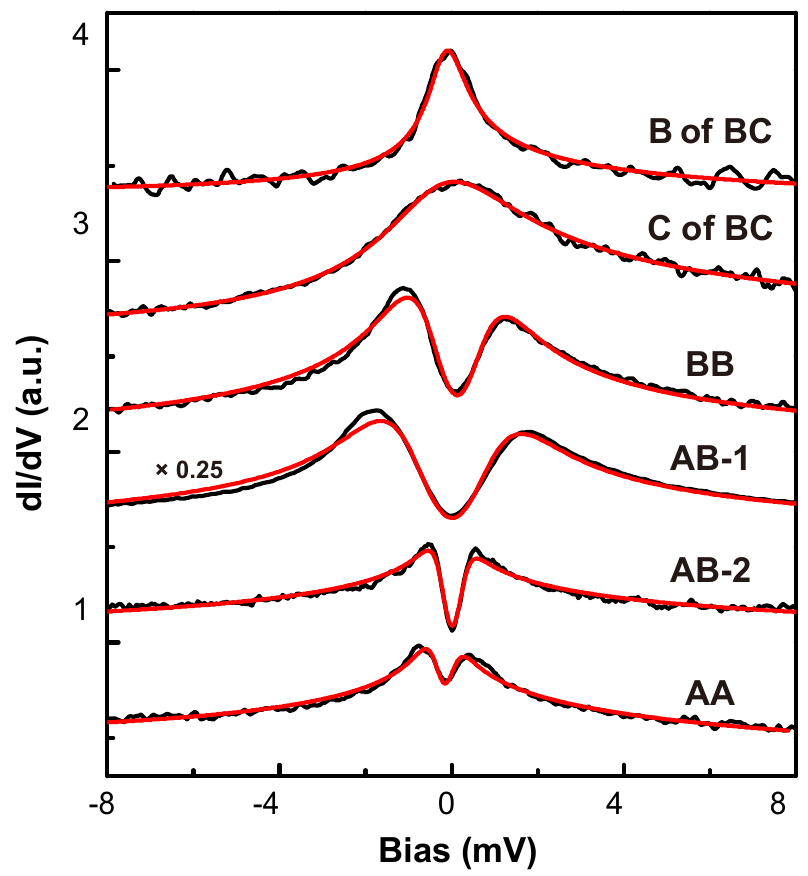}
	\caption{$dI/dV$ spectra of different ReA-radical pairs.}
	\label{FerroCoupling}
\end{figure}

\begin{table}
	\caption{Calculated spin distribution of the ReA radical.}
	\begin{ruledtabular}
		\begin{tabular}{cccc}
			\multicolumn{4}{c}{Index position}\\
			\multicolumn{4}{c}{\includegraphics[width=8.8cm]{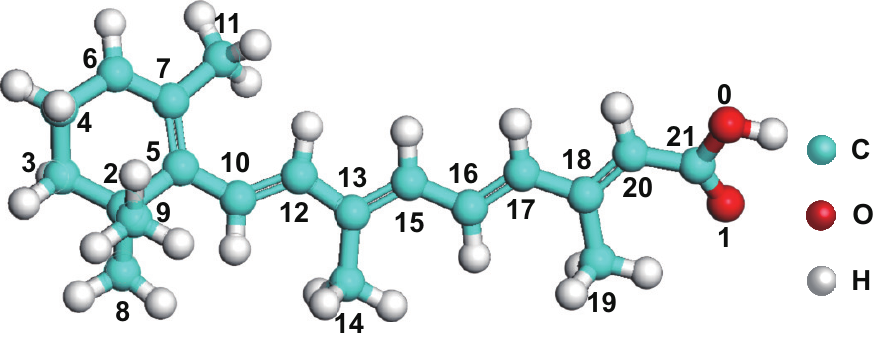}}\\
			\hline
			\multirow{2}{*}{Index}&\multirow{2}{*}{Atom}&
			\multicolumn{2}{c}{Magnetization}\\
			\cline{3-4}
			& & isolated & on Au(111)\\
			\hline\\
			0& O&	0.012895459 &	0.007916864\\
			1& O&	0.055978032&	0.034646892\\
			2& C&	-0.011141637&	-0.006848056\\
			3& C&	0.005941828&	0.003616858\\
			4& C&	-0.007849953&	-0.004792641\\
			5& C&	0.244161458&	0.15016833\\
			6& C&	0.192307588&	0.118127564\\
			7& C&	-0.067146478&	-0.041829899\\
			8& C&	0.005936133&	0.003652478\\
			9& C&	0.014077245&	0.008659227\\
			10 & C&	-0.082481319&	-0.051790944\\
			11 & C&	0.003476568&	0.002101445\\
			12 & C&	0.208806872&	0.129423872\\
			13 &C&	-0.079847063&	-0.049866339\\
			14 & C&	0.003887068&	0.002372436\\
			15 &C&	0.182255178&	0.112942854\\
			16 &C&	-0.064237883&	-0.040601864\\
			17 &C&	0.1699749 &	0.105801976\\
			18 &C&	-0.041480447&	-0.026203638\\
			19 &C&	0.002287142&	0.001454752\\
			20 &C&	0.128738247&	0.080155919\\
			21 & C&	0.001112992&	0.000224006

		\end{tabular}
	\end{ruledtabular}
\end{table}

\section{Valence Bond Theory}
Valence bond theory dates back to Pauling's work in the 1930'ies. Its main feature is the resonance diagrams with a set of double lines describing the various bonds ($\sigma$ and $\pi$). In the 1980'ies the $\pi$-bonds acquired a new interpretation in the work of Kuwajima and others (Refs. 32-35 in the main text). Here a bond line really represents a singlet pair of spins. In this approach the effective Hamiltonian is the Heisenberg spin Hamiltonian $\sum_{\langle ij\rangle}J_{ij} \hat{p}_{ij}$. Here $\hat{p}_{ij} = 1/4-\hat{S}_i\cdot \hat{S}_j$ is the operator which projects onto the singlet state of the two spins. $J_{ij}$ is negative and represents the energy gain by forming a bond between the nearest neighbor atoms labeled $i$ and $j$. A more correct Hamiltonian is the PPP-model, which is a true manybody Hamiltonian, where a H\"uckel model is combined with electron-electron repulsion. For the simplest molecule Ethylene with only two $\pi$ electrons the ground state is a variation of the Heitler-London state
\begin{equation}\label{HeitlerLondon}
	|\Psi_g\rangle = \frac{1}{\sqrt{2}} \big(|\uparrow,\downarrow\rangle-|\downarrow,\uparrow\rangle\big) + \frac{b}{\sqrt{2}}\big(|\circ,\uparrow\downarrow\rangle + |\uparrow\downarrow,\circ\rangle\big).
\end{equation}
This state is representing the true nature of the $\pi$-bond. It is a superposition of a singlet state of a pair of electrons, one on each atom, and a state containing ionic states. In Kuwajima's work the Heisenberg Hamiltonian is a projected version of the full PPP-Hamiltonian, and the exchange constants $J_{ij}$ is the bond strength.

\begin{figure}
	\centerline{\includegraphics[width=6cm]{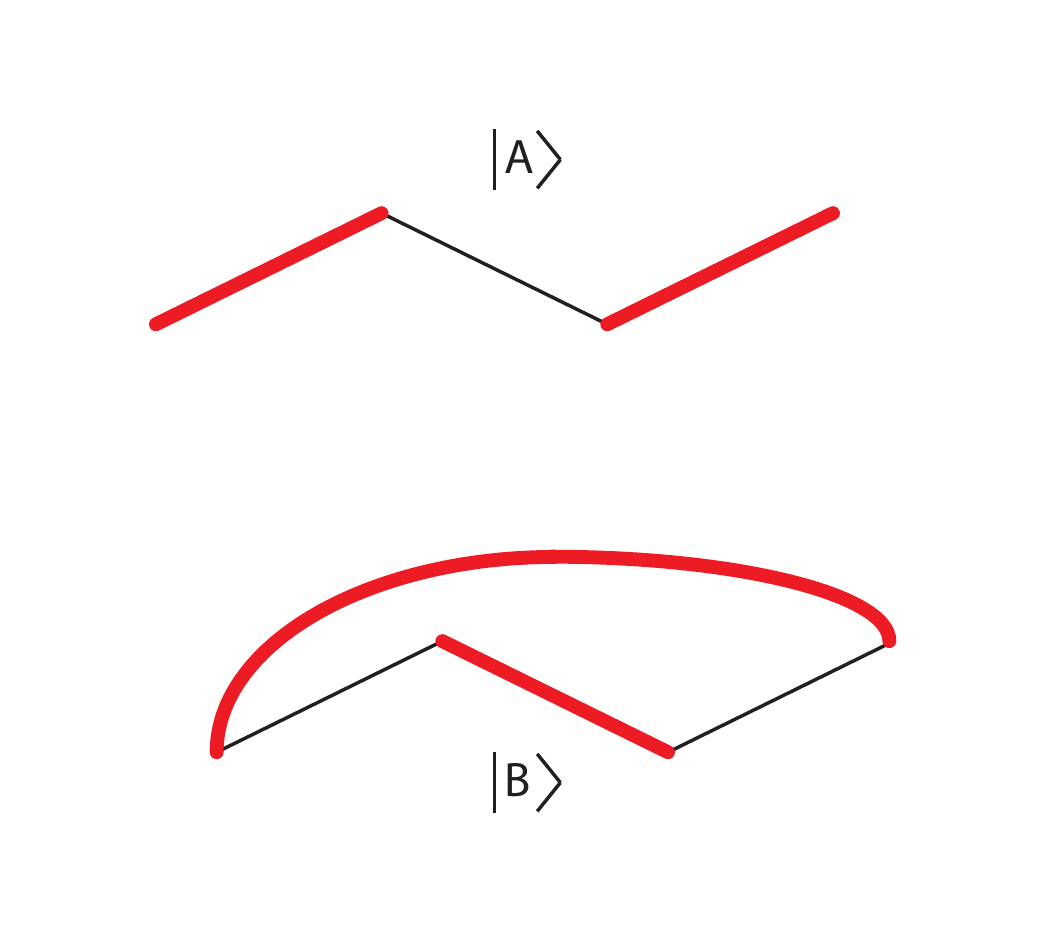}}
	\caption{\label{Fig2-5} The two valence bond states of Butadiene.}
\end{figure}

In molecules with more than two carbons, the situation is much richer. Take Butadiene as an example. Here 4 carbons form a chain. The ground state is dominated by two bonds between carbons 1 and 2 and between 3 and 4 (denoted $|A\rangle$). There is, however, also a state (or a resonance formula) where carbons 2 and 3 form a bond and carbons 1 and 4 are forming a singlet pair of electrons (denoted $|B\rangle$, see Fig.~\ref{Fig2-5}.

The non-trivial effect of the Heisenberg Hamiltonian is that the term $\hat{S}_2\cdot\hat{S}_3$ couples the two states. In fact we have
\begin{equation}\label{Coupling}
	p_{23} |A\rangle = -\frac{1}{2}|B\rangle.
\end{equation}
Using such a scheme based on the spin part of the true manybody state, one can easily generate the important resonance states, their contribution to the ground state, and the total energy of the state.

\begin{figure}[h]
	\centerline{\includegraphics[width=0.48\textwidth]{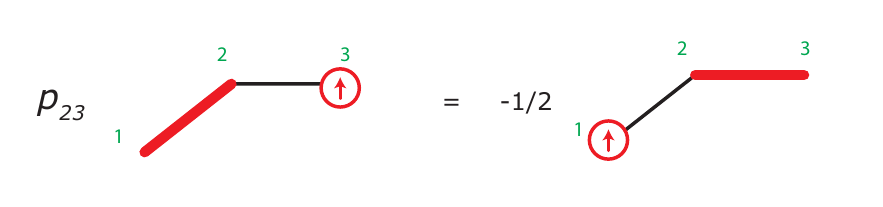}}
	\caption{\label{RadicalHop} The effect of the operator $p_{23}$ on a radical state with an unpaired spin on atom 3. Note, that the unpaired spin hops to the next nearest neighbor.}
\end{figure}

In the molecules we are studying in the present work, there are both radicals, i.e.~unpaired spins, and lone pairs in the $p_z$ orbitals. The action of the Heisenberg Hamiltonian on a radical state is shown in Fig.\ref{RadicalHop}. Note, that the radical hops to the next nearest neighbors.

\begin{figure}[h]
	\centerline{\includegraphics[width=0.48\textwidth]{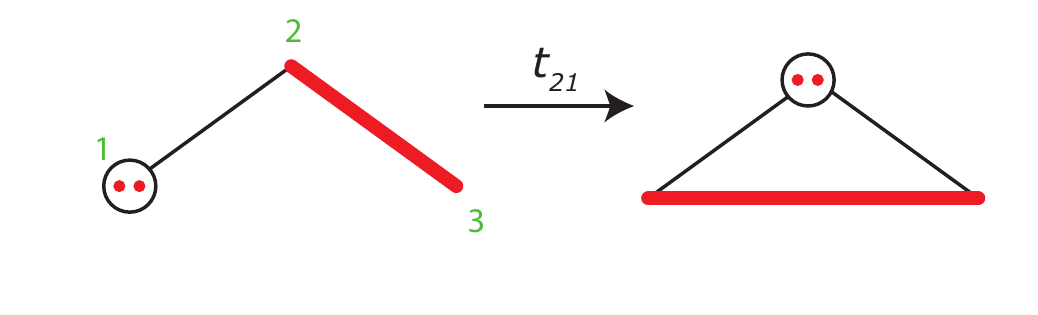}}
	\caption{\label{DoubleHop} Effect of the electron hopping operator, moving an electron from atom 1 to atom 2. The amplitude of such a hop is the H\"uckel parameter here denoted $t_{21}$. Note, that the singlet bond originally at atoms 2 and 3 moves to atoms 1 and 3.}
\end{figure}

\begin{figure}[h]
	\centerline{\includegraphics[width=0.5\textwidth]{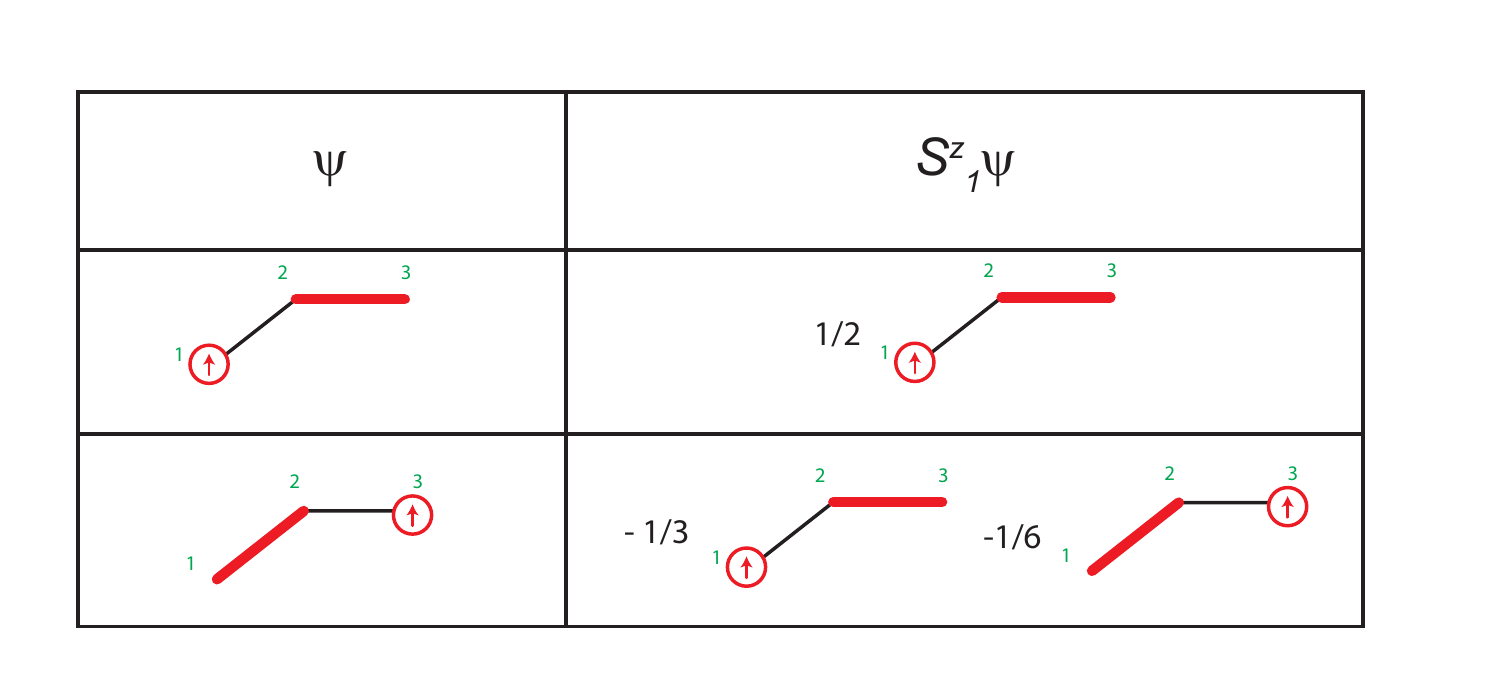}}
	\caption{\label{SpinRules} The action of the operator $S^z_1$ on various Valence Bond states.}
\end{figure}
If a molecule contains atoms with two $p_z$ electrons, as e.g.~in the OH group of the ReA molecule being studied, then rules for moving the extra $\pi$-electron needs to be specified. This rule is shown in Fig.\ref{DoubleHop}.

In the case of radical states, where the spin states involve a single unpaired electron plus a number of singlet pairs. The spin operator $S_{iz}$, which represents the $z$-component of the spin on a given carbon atom, is a non-trivial operator in the basis of spin states composed of singlet pairs and a single unpaired spin. The rules of $S_{iz}$ is show in Fig.\ref{SpinRules}.

These rules are used in the main paper, to determine the spin-density $\langle\Psi|S_{iz}|\Psi\rangle$ of the certain valence bond state $|\Psi\rangle$.

\twocolumngrid
\bibliography{ref,sm}

\end{document}